\documentclass[journal]{IEEEtran}
\usepackage{amsmath, amssymb, amsthm, bm}
\usepackage{graphicx}
\usepackage{color, colortbl} 
\usepackage[dvipsnames,svgnames,x11names]{xcolor}
\makeatother
\usepackage{tabularx, boldline, multirow}
\usepackage[boxed,ruled,vlined]{algorithm2e}
\usepackage{enumerate}
\usepackage{textcomp, gensymb}
\usepackage[caption=false,font=footnotesize]{subfig}
\usepackage{hyperref, cite}
\usepackage{url,comment}
\usepackage[normalem]{ulem}
\usepackage[shortlabels]{enumitem}
\usepackage{pgfplots}
\usepackage{tikz}
\usepackage{mathtools}
\usepackage{accents}
\usepackage{balance}
\usepackage{acronym}
\usepackage{arydshln}
\usepackage{scalerel}
\pgfplotsset{compat=1.18}

\usetikzlibrary{calc}
\makeatletter
\newcommand{\gettikzxy}[3]{%
  \tikz@scan@one@point\pgfutil@firstofone#1\relax
  \edef#2{\the\pgf@x}%
  \edef#3{\the\pgf@y}%
}

\allowdisplaybreaks
\acrodef{crb}[CRB]{Cram\'er-Rao bound}
\acrodef{adc}[ADC]{analog-to-digital converter}
\acrodef{awgn}[AWGN]{additive white Gaussian noise}
\acrodef{ap}[AP]{access point}
\acrodef{bs}[BS]{base station}
\acrodef{bp}[BP]{belief propagation}
\acrodef{1d}[1D]{one-dimensional}
\acrodef{2d}[2D]{two-dimensional}
\acrodef{3d}[3D]{three-dimensional}
\acrodef{5g}[5G]{fifth generation}
\acrodef{6g}[6G]{sixth generation}
\acrodef{ue}[UE]{user equipment}
\acrodef{nue}[NUE]{neighboring user equipment}
\acrodef{iue}[IUE]{image user equipment}
\acrodef{los}[LOS]{line-of-sight}
\acrodef{aoa}[AoA]{angle-of-arrival}
\acrodefplural{aoa}[AoAs]{angle-of-arrival}
\acrodef{aod}[AoD]{angle-of-departure}
\acrodefplural{aod}[AoDs]{angle-of-departures}
\acrodef{toa}[ToA]{time-of-arrival}
\acrodefplural{toa}[ToAs]{time-of-arrival}
\acrodef{tdoa}[TDoA]{time-difference-of-arrival}
\acrodef{ris}[RIS]{reconfigurable intelligent surface}
\acrodefplural{ris}[RISs]{reconfigurable intelligent surfaces}
\acrodef{isac}[ISAC]{integrated sensing and communications}
\acrodef{nris}[NRIS]{non-RIS}
\acrodef{cp}[CP]{cyclic prefix}
\acrodef{tx}[Tx]{transmitter}
\acrodef{rx}[Rx]{receiver}
\acrodefplural{rx}[Rxs]{receivers}
\acrodef{psd}[PSD]{power spectral density}
\acrodef{lss}[LS]{least-squares}
\acrodef{crlb}[CRLB]{Cram\'er-Rao lower bounds}
\acrodefplural{crlb}[CRLBs]{Cram\'er-Rao lower bounds}
\acrodef{rss}[RSS]{received signal strength}
\acrodef{los}[LOS]{line-of-sight}
\acrodef{nlos}[NLOS]{non line-of-sight}
\acrodef{fov}[FoV]{field-of-view}
\acrodef{dft}[DFT]{discrete Fourier transform}
\acrodef{fft}[FFT]{fast Fourier transform}
\acrodef{fmcw}[FMCW]{frequency modulated continuous wave}
\acrodef{fim}[FIM]{Fisher information matrix}
\acrodefplural{fim}[FIMs]{Fisher information matrix}
\acrodef{efim}[EFIM]{equivalent Fisher information matrix}
\acrodef{upa}[UPA]{uniform planar array}
\acrodefplural{upa}[UPAs]{uniform planar arrays}
\acrodef{peb}[PEB]{position error bound}
\acrodef{slam}[SLAM]{simultaneous localization and mapping}
\acrodef{snr}[SNR]{signal-to-noise ratio}
\acrodefplural{snr}[SNRs]{signal-to-noise ratio}
\acrodef{sre}[SRE]{smart radio environment}
\acrodefplural{sre}[SRE]{smart radio environments}
\acrodef{mpmb}[MPMB]{marginal Poisson multi-Bernoulli}
\acrodef{mb}[MB]{multi-Bernoulli}
\acrodef{pmb}[PMB]{Poisson multi-Bernoulli}
\acrodef{pmbm}[PMBM]{Poisson multi-Bernoulli mixture}
\acrodef{mimo}[MIMO]{multiple-input  multiple-output}
\acrodef{rfs}[RFS]{random finite set}
\acrodef{rfid}[RFID]{radio-frequency identification}
\acrodef{siso}[SISO]{single-input single-output}
\acrodef{miso}[MISO]{multiple-input single-output}
\acrodef{ici}[ICI]{inter-carrier interference}
\acrodef{iid}[iid]{independent and identically distributed}
\acrodef{ml}[ML]{maximum likelihood}
\acrodef{pdf}[PDF]{probability density function}
\acrodef{cdf}[CDF]{cumulative distribution function}
\acrodef{ofdm}[OFDM]{orthogonal frequency-division multiplexing}
\acrodef{qos}[QoS]{Quality of Service}
\acrodef{sp}[SP]{scattering point}
\acrodefplural{sp}[SPs]{scattering points}
\acrodef{ula}[ULA]{uniform linear array}
\acrodef{va}[VA]{virtual anchor}
\acrodefplural{va}[VAs]{virtual anchors}
\acrodef{ls}[LS]{large surface}
\acrodefplural{ls}[LSs]{large surfaces}
\acrodef{rp}[RP]{reflection point}
\acrodefplural{ls}[RPs]{reflection point}
\acrodef{eb}[EB]{error bound}
\acrodefplural{eb}[EBs]{error bounds}
\acrodef{peb}[PEB]{position error bound}
\acrodefplural{peb}[PEBs]{position error bound}
\acrodef{heb}[HEB]{heading error bound}
\acrodefplural{heb}[HEBs]{heading error bound}
\acrodef{seb}[SEB]{speed error bound}
\acrodefplural{seb}[SEBs]{speed error bound}
\acrodef{mae}[MAE]{mean absolute error}
\acrodefplural{mae}[MAEs]{mean absolute error}
\acrodef{ab}[AB]{arbitrary beam}
\acrodef{cb}[CB]{conventional beam}
\acrodef{gospa}[GOSPA]{generalized optimal subpattern assignment}
\acrodef{mae}[MAE]{mean absolute error}
\acrodef{ppp}[PPP]{Poisson point process}
\acrodef{mmwave}[mmWave]{millimeter-wave}
\acrodef{mui}[MUI]{multi-user interference}
\acrodef{if}[IF]{intermediate frequency}
\acrodef{ti}[TI]{Texas Instruments}
\acrodef{dsp}[DSP]{digital signal processing}

\newcommand{\Gtrx}{G_{\rm{trx}}}

\newcommand{\pris}{\alpha_{\rm{RIS}}}
\newcommand{\Cor}[1]{{\color{black}{#1}}} 

\begin{document}
\bstctlcite{IEEEexample:BSTcontrol}
\title{RIS-Enabled Self-Localization with FMCW Radar}
\author{
\IEEEauthorblockN{
Hyowon~Kim, \IEEEmembership{Member, IEEE},
Navid~Amani,
Musa~Furkan~Keskin, \IEEEmembership{Member, IEEE}, \\
Zhongxia Simon He, \IEEEmembership{Senior Member, IEEE},
Jorge Gil, \\
Gonzalo-Seco Granados, \IEEEmembership{Fellow, IEEE},
and Henk Wymeersch, \IEEEmembership{Fellow, IEEE}}
\thanks{H. Kim is with the Department of Electronics Engineering, Chungnam National University,  Daejeon, South Korea (email: hyowon.kim@cnu.ac.kr).
M. F. Keskin, N. Amani and H. Wymeersch are with the Department of Electrical Engineering, Chalmers University of Technology, Gothenburg, Sweden (email: furkan, anavid, henkw@chalmers.se).
Z. S. He is with Chalmers Industriteknik, Gothenburg, Sweden (email: simon.he@chalmersindustriteknik.se).
J. Gil is with the Department of Architecture and Civil Engineering, Chalmers University of Technology, Gothenburg, Sweden (email: jorge.gil@chalmers.se).
G. Seco Granados is with the Universitat Autonoma de Barcelona, Barcelona, Spain (email: gonzalo.seco@uab.cat)
}
\thanks{This work was supported by the Chalmers Transport Area of Advance, the Swedish Research Council grant 2022-03007 and 2023-05184.}
}
\maketitle

\begin{abstract}
    In the upcoming vehicular networks, \acp{ris} are considered as a key enabler of user self-localization without the intervention of the \acp{ap}.
    In this paper, we investigate the feasibility of \ac{ris}-enabled self-localization with no \acp{ap}.
    We first develop a \ac{dsp} unit for estimating the geometric parameters such as the angle, distance, and velocity and for \ac{ris}-enabled self-localization.
    Second, we set up an experimental testbed consisting of a Texas Instrument \ac{fmcw} radar for the user and SilversIMA module for the \ac{ris}.
    Our results confirm the validity of the developed DSP unit and demonstrate the feasibility of RIS-enabled self-localization.
\end{abstract}

\begin{IEEEkeywords}
    Channel parameters estimation, frequency modulated continuous wave radar, localization, reconfigurable intelligent surface. 
\end{IEEEkeywords}

\IEEEpeerreviewmaketitle
\vspace{-7mm}
\section{Introduction}\label{sec:Introduction}
Navigation, especially in unknown areas without access to global navigation satellite systems (GNSS), presents a significant challenge \cite{chiang2019seamless,GNSS_Lidar_SLAM_2021}. Simultaneous localization and mapping (SLAM) offers a promising solution by allowing a system to map and localize itself simultaneously in real-time \cite{Durrant2006SLAM1,Durrant2006SLAM2}. However, traditional SLAM systems heavily rely on optical sensors like cameras or LiDAR, which are limited in vision-denied environments—conditions such as smoke-filled rooms, darkness, or areas with poor visibility \cite{radioSLAM_review_2023,UWB_SLAM,santos2015sensor}. These are conditions commonly encountered by first responders during rescue missions or in industrial environments \cite{santos2015sensor}. 

To address these limitations, frequency modulated continuous wave (FMCW) radar has emerged as a viable alternative \cite{fmcw_survey_2022,fmcw_SPM_2017}. Unlike optical sensors, FMCW radar can operate reliably under various weather conditions, including fog, rain, or dust \cite{Canan_FMCW_SPM2020}. It has already been successfully adopted for autonomous vehicles and robotics \cite{radarSLAM_2021}, making it a potential solution for SLAM in harsh or low-visibility environments \cite{slam_lidar_exp_2018}.
Radar odometry enhances SLAM by analyzing features in the environment, such as static objects like walls, through consecutive radar scans. By 
matching environmental features across several radar scans
radar odometry can estimate the movement of a vehicle or robot \cite{radarOdometry_2024}. 

Rather than relying solely on natural features for localization, we propose to leverage reconfigurable intelligent surfaces (RISs) as artificial landmarks to enhance \textit{radar self-localization} in SLAM applications. Although RISs have been extensively studied in the context of radar-like sensing of \textit{passive} objects in monostatic \cite{Zhang_MetaRadar_2022,RIS_sensing_JSAC_2022,NLOS_RIS_sensing_CRB_TSP_2023,ris_radar_clutter_2024,RIS_multiTarget_TCOM_2022,RIS_mono_Hyowon_2023}, bistatic \cite{foundations_RIS_radar_TSP_2022,NLOS_RIS_Sensing_2023,ISAC_RIS_WCM_2023} and multistatic \cite{wei_target_2021} configurations, their use in self-localization of an \textit{active} transmitting device (i.e., a full-duplex radar) remains relatively under-explored, e.g., \cite{Kamran_RISloc_ICC2022,RIS_AP_Free_SLAM_Hyowon_2024}. RISs as artificial landmarks can be strategically placed in the environment, offering consistent and reliable reference points for radar systems \cite{Keykhosravi2022infeasible,Kamran_RISloc_ICC2022}. Moreover, such RIS-assisted self-localization solution obviates the need for additional costly infrastructure such as access points (APs) or base stations (BSs) \cite{RIS_WCM_6G_2021,Hui_RISLocSens_2023} and provides a cost-effective strategy to implement radar-based SLAM \cite{RIS_AP_Free_SLAM_Hyowon_2024}.
While RIS-assisted self-localization has been theoretically investigated in \cite{Kamran_RISloc_ICC2022,RIS_AP_Free_SLAM_Hyowon_2024}, its feasibility has not been demonstrated in a real-world experimental setup.

\begin{figure}
    \centering
    \includegraphics[width=0.99\linewidth]{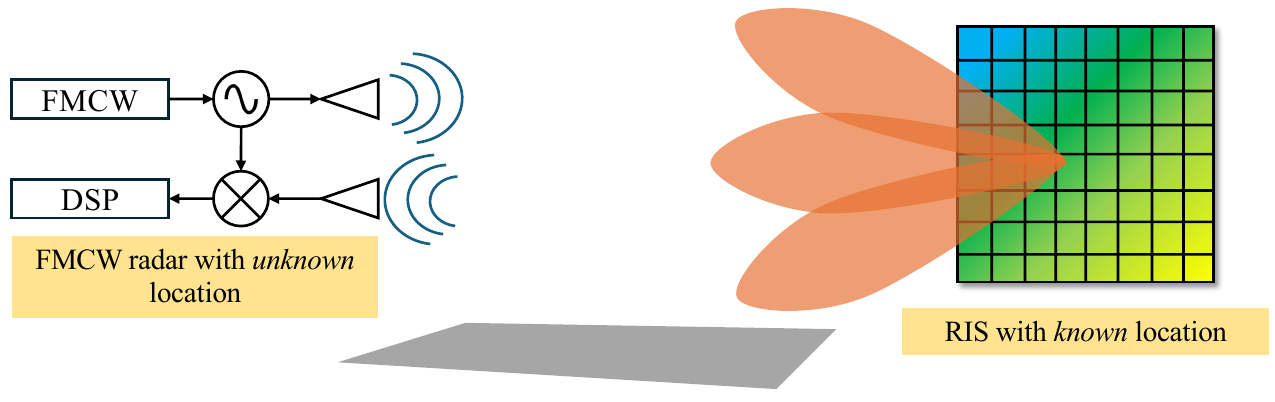}
    \caption{Considered scenario where an FMCW radar is localized using an RIS. }
    \label{fig:scenario}
\end{figure}

In this paper, we aim to bridge this gap by designing an experimental setup for RIS-aided self-localization of a monostatic FMCW radar (shown in Fig.~\ref{fig:scenario}), accompanied by algorithms for RIS beam scanning, distance and AOD estimation for localization, and a simulator calibrated to match experimental results and evaluate performance under a wide range of realistic conditions.  
Our specific contributions are: \textit{(i)} We develop a method tailored for RIS-aided self-localization, allowing real-time estimation of critical geometric parameters such as angle of departure, distance, and velocity in a monostatic radar configuration; \textit{(ii)} We construct a comprehensive experimental testbed, incorporating an FMCW radar from Texas Instruments~\cite{rao2017introduction} 
to emulate the UE and a transceiver from SiversIMA~\cite{Silvers6002} 
to emulate the RIS, to validate the proposed self-localization approach under realistic conditions; \textit{(iii)} We evaluate the feasibility and accuracy of RIS-enabled self-localization through experimental data analysis, and develop a simulator calibrated with this data to investigate system performance under various scenarios, including changes in transmit power, RIS sweep step size, and RIS array dimensions.

\vspace{-3mm}
\section{System Model}
\vspace{-1mm}
    We consider a full-duplex \ac{ue} and a passive \ac{ris}.
    The \ac{ris} is equipped with an \ac{upa} composed of $N_\text{RIS} = N_\text{RIS}^\text{az} \times N_\text{RIS}^\text{el}$ radiating elements,
    which lies on the $xz$ plane,
    and the \ac{ue} is equipped with a single antenna. Both UE and RIS are static.
    %
    The \ac{ue} location is denoted by $\mathbf{x}_\text{UE}$. 
    The geometric center of elements at the \ac{ris} is denoted by $\mathbf{x}_\text{RIS}$ 
    and the location of the $n$-th element is denoted by $\mathbf{x}_\text{RIS}^n$.
    


    The \ac{ue} transmits the chirp signals, and the \ac{ris} reflects the signals by controlling the \ac{ris} phases, performing a sweeping action across the angular range.  
    The other surrounding objects reflect the transmitted signals.
    Then, the \ac{ue} receives multipath due to reflection or scattering in the propagation environment consisting of the RIS and other objects.
    The \ac{adc} sample, chirp, 
    \ac{ris} sweep angle, and signal path are respectively indexed by $n$, $k$, 
    $m$, and $l$.
    The beat signal at the \ac{if} for the $m$-th \ac{ris} sweep angle is denoted by $\mathbf{Y}_m$ whose element $[\mathbf{Y}_{m}]_{n,k}$ indicates the signal on the $n$-th sample for each $k$-th chirp, given by~\cite{Canan_FMCW_SPM2020}    
\begin{align}
     [\mathbf{Y}_{m}&]_{n,k} =\gamma_0 e^{-j2\pi(S\tau_0 + f_c \nu_0)nT_s}e^{-j2\pi f_c \nu_0 k T} \mathbf{a}^\top(\bm{\theta})\bm{\Omega}_{m}\mathbf{a}(\bm{\theta}) \notag \\
    &+ \sum_{l=1}^L \gamma_l e^{-j2\pi(S\tau_l + f_c \nu_l)n T_s}e^{-j2\pi f_c \nu_l k T}  + 
    w_{n,k,m},
    \label{eq:beat_signal}
\end{align}
    where the index $l=0$ indicates the reflected signals from the \ac{ris}  and $l>0$ for the 
    reflected signals from the other objects, the total number of signal paths is $L+1$, $\gamma_l$ is the complex path gain,
    $\mathbf{a}(\bm{\theta}) \triangleq \exp(j {\mathbf{X}}_\text{RIS}^\top \mathbf{g}(\bm{\theta}))$ denotes the array vector at the \ac{ris} with $\bm{\theta} \triangleq [\theta^\text{az},\theta^\text{el}]^\top$ being the \ac{aoa} and \ac{aod} at the \ac{ris},\footnote{In monostatic systems, the \ac{aoa} at an object is the same as the \ac{aod}.} 
    $\mathbf{g}(\bm{\theta}) \triangleq \frac{2\pi}{\lambda}[\cos(\theta^\text{az})\sin(\theta^\text{el}), \sin(\theta^\text{az})\sin(\theta^\text{el}), \cos(\theta^\text{el})]^\top$ is the wavenumber vector, $\lambda = c/f_c$ is the wavelength, $c$ is the speed of light, $f_c$ is the carrier frequency, $\mathbf{X}_\text{RIS}\triangleq [\mathbf{x}_\text{RIS}^1,\dots,\mathbf{x}_\text{RIS}^{N_\text{RIS}}] $,
    $\bm{\Omega}_{m}=\text{diag}(\bm{\omega}_{m})$ is the \ac{ris} phase matrix, and $\bm{\omega}_{m}\triangleq \exp(-2j \mathbf{X}_\text{RIS}^\top\mathbf{g}(\bm{\phi}_m))$ is the RIS phase profile vector whose elements lie on the unit circle with $\bm{\phi}_m\triangleq [\phi_m^\text{az},\phi_m^\text{el}]^\top$ being the parameter vector corresponding to the $m$-th sweep angle at the \ac{ris}. During $m$-th sweep angle, i.e., seen as a frame, $K$ chirps are transmitted, and each chirp consists of $N$ \ac{adc} samples.
    Furthermore, $S=B/T$ denotes the chirp slope, $B$ is the sweep bandwidth, 
    $T$ is the chirp duration, 
    $T_s$ is the sample duration, 
    $\tau_l=2d_l/c$ is the round-trip time delay, $d_l$ is the true distance between the \ac{ue} and object producing the $l$-th reflection, whose radial velocity is denoted by $v_l$, and hence $\nu_l=2v_l/c$ is the corresponding Doppler shift,
     \Cor{$w_{n,k,m} \sim \mathcal{CN}(0,\sigma_\mathrm{N}^2)$ is the complex Gaussian noise, and $\sigma_\mathrm{N}^2$ is the signal noise covariance.}
\section{RIS-Enabled Self-Localization}
\label{sec:SelfLoc}
This section describes the signal processing at the FMCW receiver. 
\vspace{-3mm}
\subsection{Delay and Doppler Spectrum}

    To jointly estimate the distance and radial velocity, we apply the \ac{2d} \ac{dft} to the beat signal~\eqref{eq:beat_signal} for the sample and chirp dimensions, generating the \ac{fmcw} delay-Doppler spectrum as described below.
    We first utilize windowing on the signals 
\begin{align}
   \bar{\mathbf{Y}}_{m} = \mathbf{w}_N \mathbf{w}_K^\top \odot  \mathbf{Y}_{m} \,,
\end{align}
    where $\mathbf{w}_A$ is the window vector with $A$ samples and $\odot$ denotes the Hadamard (element-wise) product. 
    To increase the estimation accuracy in \ac{2d} \ac{dft}, we adopt the zero padding  
\begin{align}
    \mathbf{Y}'_{m}=
\begin{bmatrix}
    \bar{\mathbf{Y}}_{m} & \bm{0}_{N \times (K_\text{DFT}-K)}\\
    \bm{0}_{(N_\text{DFT}-N) \times K} & \bm{0}_{(N_\text{DFT}-N) \times (K_\text{DFT}-K)}
\end{bmatrix},
\end{align}
    where $N_\text{DFT}$ and $K_\text{DFT}$ are respectively the number of \ac{dft} samples for \ac{adc} samples and chirps.  
    For the different \ac{ris} beam sweep angles, i.e., $m=1,\dots,M$, we generate the delay-Doppler map with the \ac{2d} \ac{dft} 
\begin{align}
    z_{m}({\tau},{\nu}) = \sum_{k=0}^{K_\text{DFT}-1} \sum_{n=0}^{N_\text{DFT}-1}  [\mathbf{Y}'_{m}]_{n,k} e^{j2\pi S {\tau} n T_s} e^{j2 \pi f_c {\nu} k T}.
    \label{eq:2DDFT}
\end{align}

\vspace{-4mm}
\subsection{Geometric Parameter Estimation}
\label{sec:GeoParaEst}
    We estimate the following geometric parameters: \ac{aod} from the \ac{ris}, distance between the \ac{ue} and \ac{ris}, and Doppler shift at the \ac{ue}.
    The geometric parameters are estimated as follows.
    %
    First, 
    we determine the delay and Doppler pairs for the different \ac{ris} beam sweep angles, i.e., $m=1,\dots,M$,
\begin{align}
    ({\tau}_{m},{\nu}_{m}) = \operatorname*{argmax}_{{\tau},{\nu}} \, \lvert z_{m}({\tau},{\nu}) \rvert^2,\label{eq:argmaxmap}
\end{align}
%
    Second, we estimate the \ac{aod} at the \ac{ris} by \ac{ris} beam sweeping.
    The received power for each beam is determined by $ \hat{m}  = \operatorname*{argmax}_{m} \,  P_{\text{ave},m}$, where
\begin{align}   
       P_{\text{ave},m} & = \frac{1}{2\Delta}\int_{\tau_m-\Delta}^{\tau_m+\Delta} \lvert z_{m}({\tau},{\nu}_m) \rvert^2 \rm{d} \tau,
       \label{eq:P_ave}
\end{align}
where $\Delta$ is a design parameter. 
    The estimated \ac{aod} $\hat{\bm{\theta}} \triangleq [\hat{\theta}^\text{az}, \hat{\theta}^\text{el}]^\top$ is the $\hat{m}$-th \ac{ris} beam $\hat{\bm{\theta}} = \bm{\phi}_{\hat{m}}$.
%
    Finally, the distance and Doppler are estimated by selecting $\hat{m}$-th delay and Doppler pair, i.e., 
    $(\hat{\tau},\hat{\nu}) = ({\tau}_{\hat{m}},{\nu}_{\hat{m}} )$,
    and the estimated distance and Doppler are determined by $\hat{d} = \hat{\tau} c /2$ and $\hat{v} = \hat{\nu} c /2$, respectively. 
    
  \vspace{-4mm}  
\subsection{Localization}
    We estimate the \ac{ue} location by exploiting the estimated distance $\hat{d}$ and \ac{ris} \ac{aod} $\hat{\bm{\theta}}$.
    The estimated \ac{ue} location is
\begin{align}
    \mathbf{x}_\text{UE} = \mathbf{x}_\text{RIS} + \begin{bmatrix}
        \hat{d}\cos (\hat{\theta}^{\text{az}}) \sin (\hat{\theta}^{\text{el}})\\
        \hat{d}\sin (\hat{\theta}^{\text{az}}) \sin (\hat{\theta}^{\text{el}})\\
        \hat{d} \cos (\hat{\theta}^{\text{el}})
    \end{bmatrix}.
\end{align}



\section{Experimental and Simulation Setup}
\subsection{Experimental Testbed}
\begin{figure}
\begin{centering}
	\subfloat[\label{Fig:ExpEnv}]
 {\includegraphics[width=.3\columnwidth]{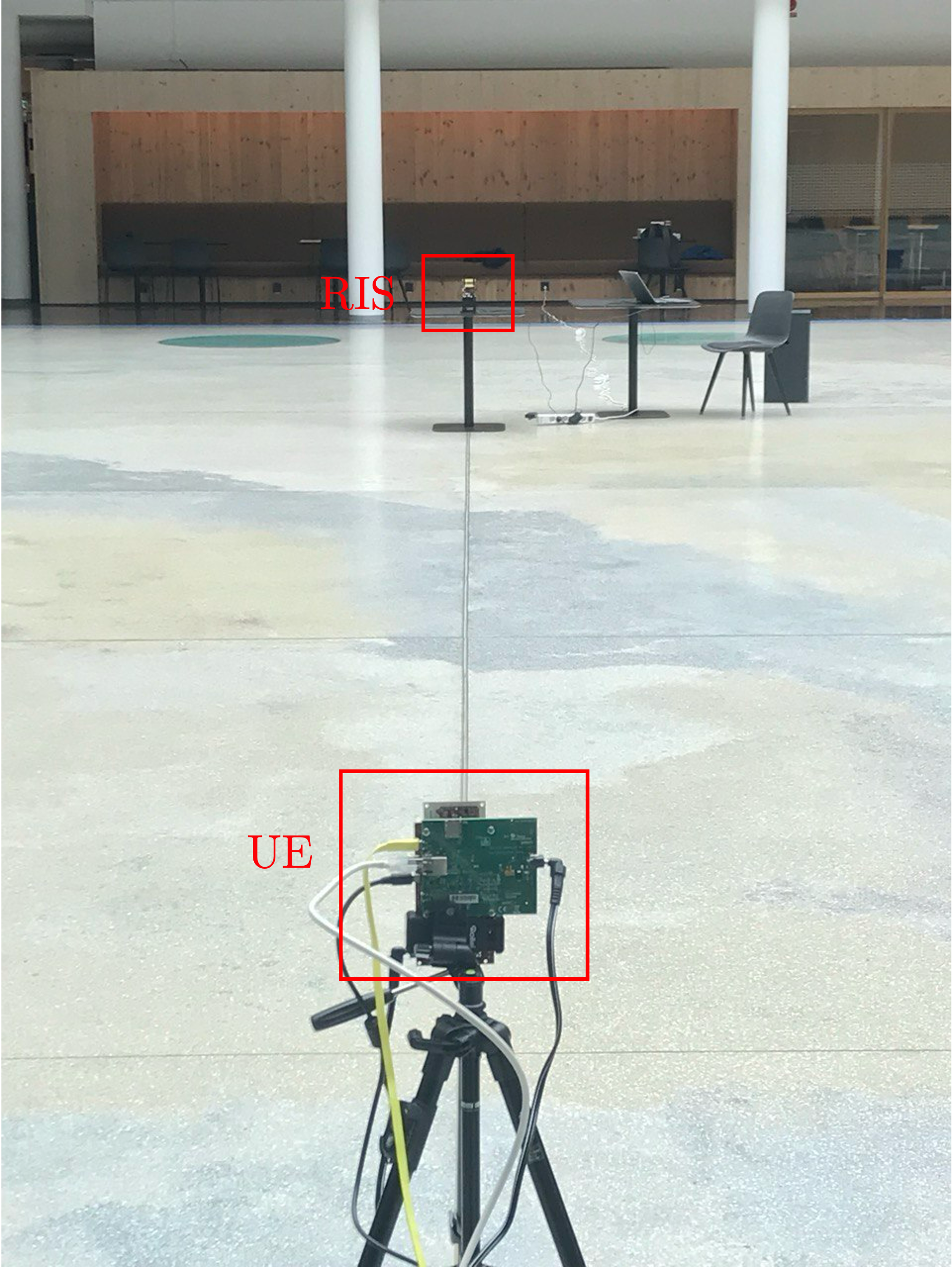}}~~
	\subfloat[\label{Fig:UE}]{\includegraphics[width=.3\columnwidth]{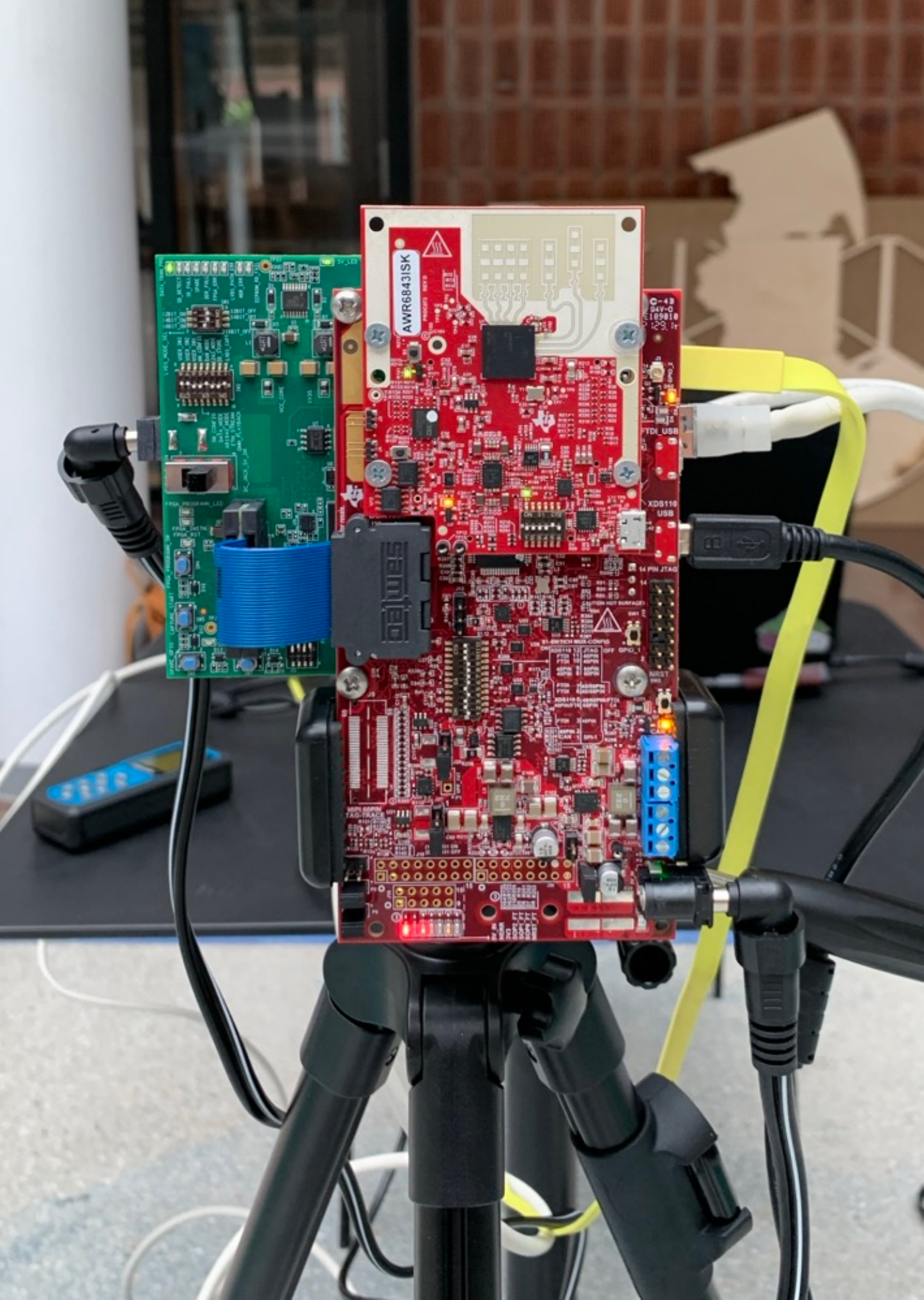}}~~
	\subfloat[\label{Fig:RIS}]{\includegraphics[width=.3\columnwidth]{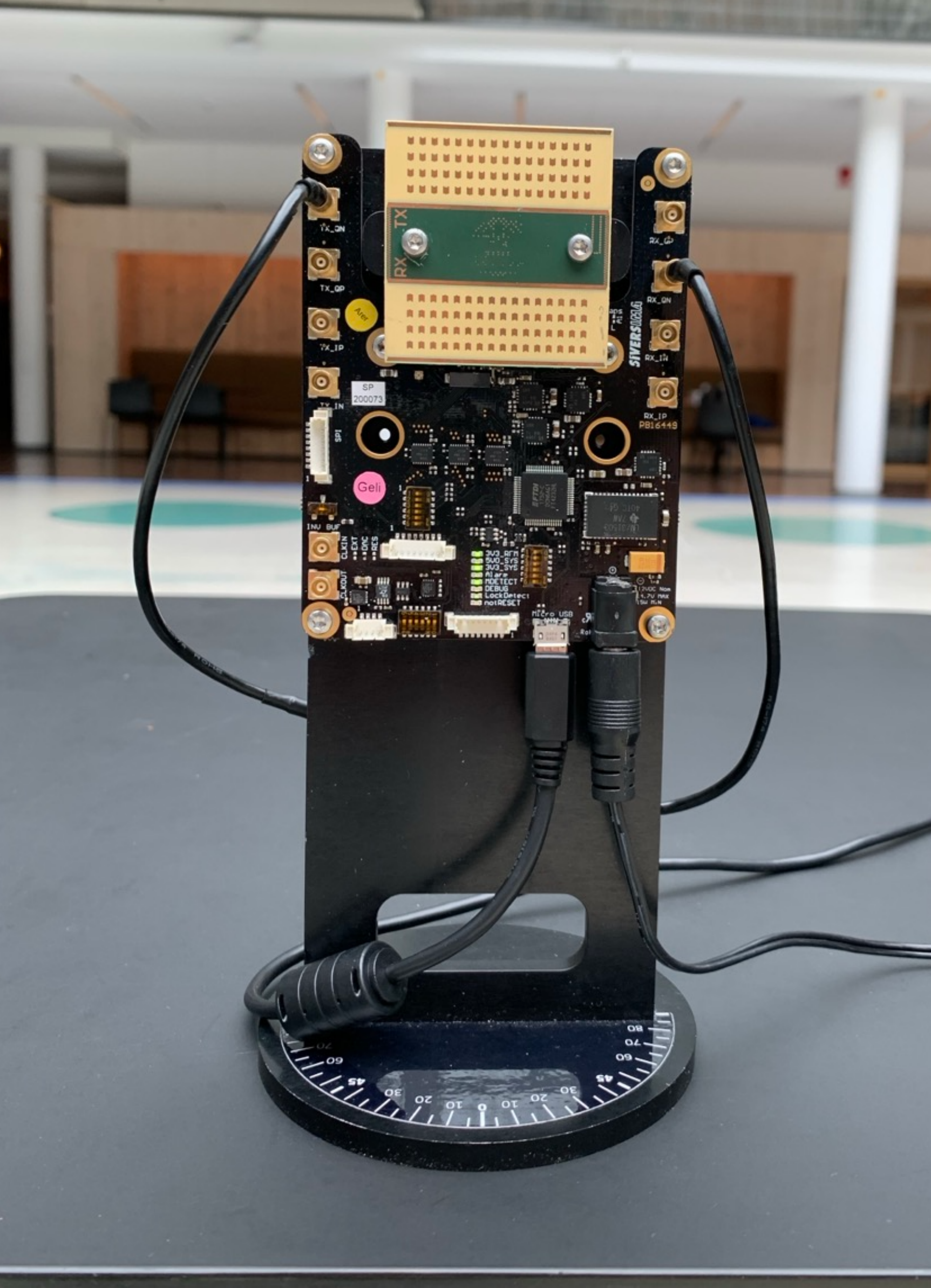}}
	\caption{Experiment environment, where UE and  RIS are deployed: (a) displacement of UE and RIS; (b) UE; and (c) RIS.}
	\label{Fig:Experiment}
	\par
\end{centering}
\end{figure}

    Our experimental testbed for \ac{ris}-enabled self-localization is depicted in Fig.~\ref{Fig:Experiment}. 
    We deploy the \ac{ue} and \ac{ris} in a large hall, separated by a clear \ac{los} path at a distance of approximately 14 meters, as shown in Fig.~\ref{Fig:ExpEnv}. 
    Both are facing each other and located at the same height, indicating that the \ac{aod} at the \ac{ris} is set to $\bm{\theta}=[0,0]^\top$.
    We emulate the \ac{ue} by the \ac{fmcw} radar evaluation kit (AWR6843ISK) from \ac{ti}~(see, Fig.~\ref{Fig:UE}), and \ac{ris} by the mmWave wireless communication module from SiversIMA~(see, Fig.~\ref{Fig:RIS}) to mimic RIS beamforming capabilities. 
    Both operate at the frequency of 60 GHz. To complement the 
    radar module at the \ac{ue}, we deploy the DCA1000EVM for real-time data capture and streaming. This configuration allows us to seamlessly collect radar data for further processing. 
    We developed an in-house \ac{dsp} unit using MATLAB, which processes the radar-captured data to extract the relevant experimental results.
    The SiversIMA module is equipped with separate transmit~(Tx) and receive~(Rx) channels, where each consists of $16 \times 4$ antenna elements, as illustrated in Fig.~\ref{Fig:RIS}, which allows it to receive radar chirps from the radar module and transmit them back in a specific direction.
    The beam-steering functionality of the \ac{ris} enables directional re-transmissions within an angle range of -45 to +45 degrees relative to its broadside, making it highly adaptable for our purposes.
    During the experiment, the \ac{ue} continuously transmits \ac{fmcw} chirps, which are received by the Rx channels of the RIS. These signals are then re-transmitted by the RIS at varying scan angles, covering the entire RIS angular range $\phi_m^\text{az}$ from -45 to 45 degrees in 1.5-degree increments\footnote{To emulate the RIS beam-steering process with separate Tx and Rx arrays, the Tx beamforming vector at the $m$-th sweep angle is $\bm{\omega}^{\text{Tx}}_{m} = \exp(-j \mathbf{X}_\text{Tx}^\top\mathbf{g}(\bm{\phi}_m))$, defined with the RIS phase matrix of~\eqref{eq:beat_signal}. Since in our experiment the RIS is located at the boresight of the UE for maximum reception, its receive beam is always directed toward the UE and does not change with the Tx beam scanning. Consequently, the Rx beamforming vector is $\bm{\omega}^{\text{Rx}} = \exp(-j \mathbf{X}_\text{Rx}^\top\mathbf{g}(\bm{\mathbf{\theta}}))$. Due to identical arrays, we have $\mathbf{X}_\text{Tx} = \mathbf{X}_\text{Rx} $.} 
    and $\phi_m^\text{el}=0^\circ$. 
    While this process takes place, the UE captures samples of the reflected signals at each scan angle.
    This experimental method allows us to measure the average
    power
    at the radar for different beam-steering angles on the RIS side, $P_{\text{ave},m}$ of~\eqref{eq:P_ave} in dBm.
    By systematically varying the angles and recording the response of the radar, we estimate the RIS \ac{aod} that maximizes the average power $P_{\text{ave},m}$. 
    Treating the RIS as an artificial landmark with a known location allows the radar to accurately self-localize, while simultaneously generating environmental images for mapping purposes. 

\subsection{Simulation Setup}
    To validate the designed signal model of~\eqref{eq:beat_signal} and the proposed RIS-enabled self-localization method in Sec.~\ref{sec:SelfLoc}, we construct a simulator of our experimental testbed in the MATLAB environment. The goal is to harness the experimental results for calibrating the simulator, which are then employed to investigate performance under a rich set of conditions that cannot be tested with experiments. Instead of real-time signal capturing and streaming, we artificially generate the noisy signal data that follows the received signal model~\eqref{eq:beat_signal}.
    The squared amplitude of complex path gain for the targets (i.e., $l>0$) is  $ \lvert \gamma_l \rvert^2 = 
        P{\Gtrx S_\text{RCS}^l \lambda^2}/({(4\pi)^3 d_l^4})$, where
    $P$ is the transmit power, $\Gtrx$ is the combined transmit and receive antenna gain, $S_\text{RCS}^l$ is the radar cross section~(RCS) of the objects that producing the signal reflection. 
    As to the complex gain of the \ac{ris} loopback path~($l=0$), it involves the impact of both amplification factors due to the active nature of the RIS \cite{activeRIS_2023} and the losses incurred during loopback from receive to transmit path in the active RIS: $ \zeta = L_{\text{loss}} \pris$, 
    where $\pris$ is the amplification factor common to all the RIS elements \cite{activeRIS_2023,activeRIS_radar_2023} and $L_{\text{loss}}$ is the loopback loss, both representing unitless quantities. Hence, the squared amplitude of the complex gain of the RIS loopback path can be expressed as \cite{activeRIS_radar_2022,activeRIS_radar_2023} 
\begin{align}
   \lvert \gamma_0 \rvert^2 = P \Gtrx \lvert \gamma_\text{UR} \rvert^2 \zeta \lvert \gamma_\text{RU} \rvert^2 \,,
\end{align}
where $\gamma_\text{UR}$ and $\gamma_\text{RU}$ denote the gains of the \ac{ue}-RIS path and the RIS-\ac{ue} radar path, respectively. We have 
\cite{activeRIS_radar_2023}  $  \lvert \gamma_\text{UR} \rvert^2 = \lvert \gamma_\text{RU} \rvert^2 = {\lambda^2}/{(4 \pi d_0)^2}$, 
where $d_0$ is the distance between the \ac{ue} and the \ac{ris}. In the computation of average power~\eqref{eq:P_ave}, the design parameter $\Delta$ of \eqref{eq:P_ave} is set to 0.33~ns, the time delay corresponding to the distance 0.1~m.
We utilize the Hann window function, and the $a$-th sample of the Hann window vector is $\mathbf{w}_A(a) = \sin^2(a\pi/A)$, where $A$ is the number of samples. To consider the re-transmission process at the RIS, we adopt the loop-back delay $\tau_\text{RB}$, and the estimated distance is computed by $\hat{d} = (\hat{\tau} - \tau_\text{RB})c/2$.
The simulation parameters are summarized in Table~\ref{tab:SimulPara}. The results are averaged over 1000 Monte Carlo simulation runs.

\begin{table}
\centering
\caption{The Simulation Parameters
Used in the Performance Evaluations.}
\label{tab:SimulPara}
\begin{tabular}{ll}
\hlineB{3}
    Parameter & Value 
    \\ \hline \hline
    \ac{ris} array size & $N_\mathrm{RIS} = 64$~($16\times4$) 
    \\
    Transmission power & $P = 20$ dBm \\
    Combined Tx and Rx antenna gain 
    & $\Gtrx=4.7712$ dBi
    \\
    RCS of the targets (for $l>0$)& 
    $S_\text{RCS}^l=19$ m$^2$\\
    Amplification and loss factor 
    & $\zeta=45.532$~dB \\
    Number of chirps & $K=128$\\
    Number of ADC samples & $N=600$\\
    Chirp duration & $T = 50$ ms \\
    Sweep bandwidth & $B=3.4345$ GHz \\
    \Cor{IF bandwidth} & \Cor{$B_\text{IF}=10$ MHz} \\
    Carrier frequency & $f_c = 60$ GHz \\
    Wavelength & $\lambda=0.005$ m \\
    \Cor{Signal noise covariance} & \Cor{$\sigma_\mathrm{N}^2 = -63.64$ dBm} \\
    Loop-back delay at the RIS & $\tau_\text{RB} = 1.78$~ns\\
    Number of DFT samples & $N_\text{DFT}=1199$, $K_\text{DFT}=4793$ \\
    UE position & $\mathbf{x}_\text{UE} = [0,0]^\top$\\
    RIS position & $\mathbf{x}_\text{RIS}=[0, 13.38]^\top$.\\
	\hlineB{3}
\end{tabular}
\end{table}  

\begin{figure}
\begin{centering}
    {\includegraphics[width=1\columnwidth]{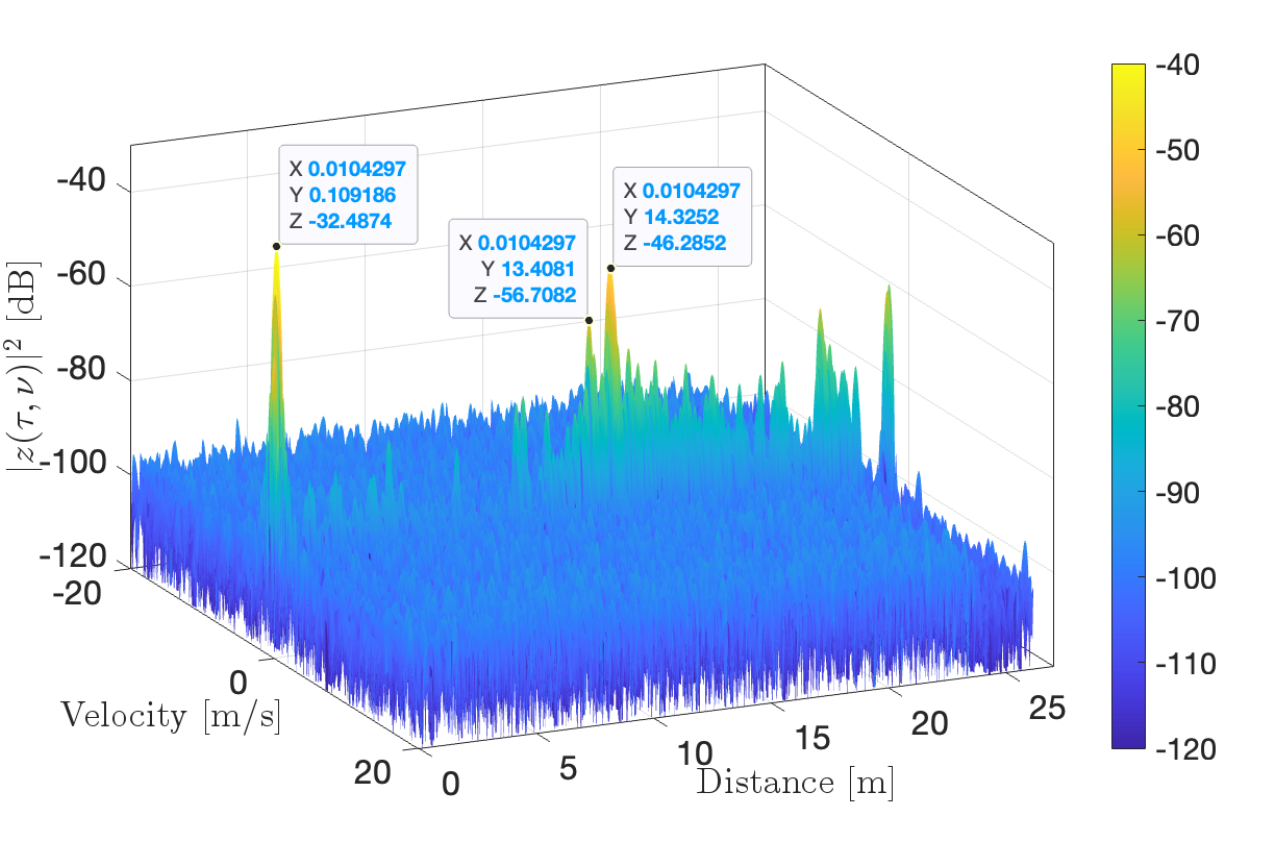}}
    \par
    \caption{Three largest peaks in 2D distance and velocity spectrum with the experiment data, obtained by~\eqref{eq:2DDFT} at the $\hat{m}$-th RIS sweep angle corresponding to $\bm{\phi}=[0\degree,0\degree]^\top$.}
    \label{Fig:RDM}
\end{centering}
\vspace{-0mm}
\end{figure}
\vspace{-2mm}
\section{Results and Discussions}
\subsection{Experimental Results}

\begin{figure}
\begin{centering}
    {\includegraphics[width=1\columnwidth]{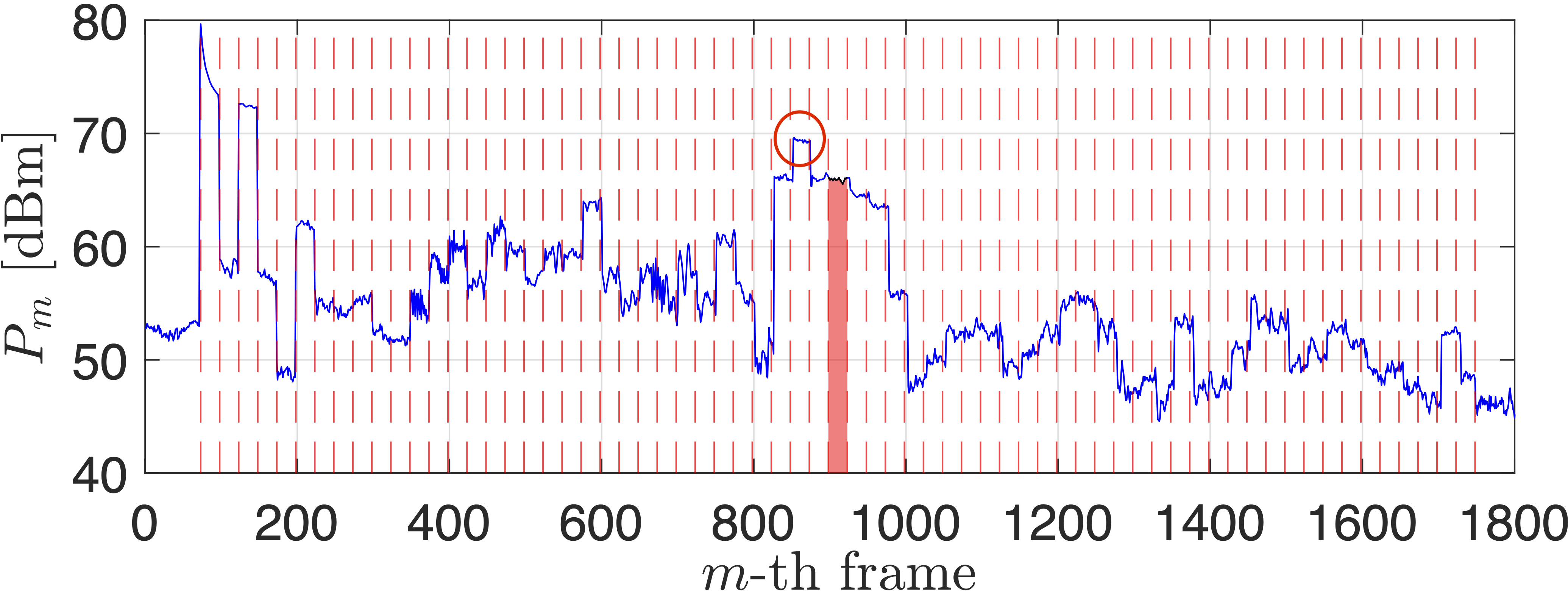}}
    \par
    \caption{Example of the  AoD estimation ($\hat{\bm{\phi}}=[-3\degree,0\degree]^\top$ and $\bm{\phi}=[0\degree,0\degree]^\top$  corresponding to the frames marked by the red circle (estimates) and highlighted by the red bar (true), respectively.).}
    \label{Fig:AoD}
\end{centering}
\end{figure}

\begin{figure}
\begin{centering}
    {\includegraphics[width=0.98\columnwidth]{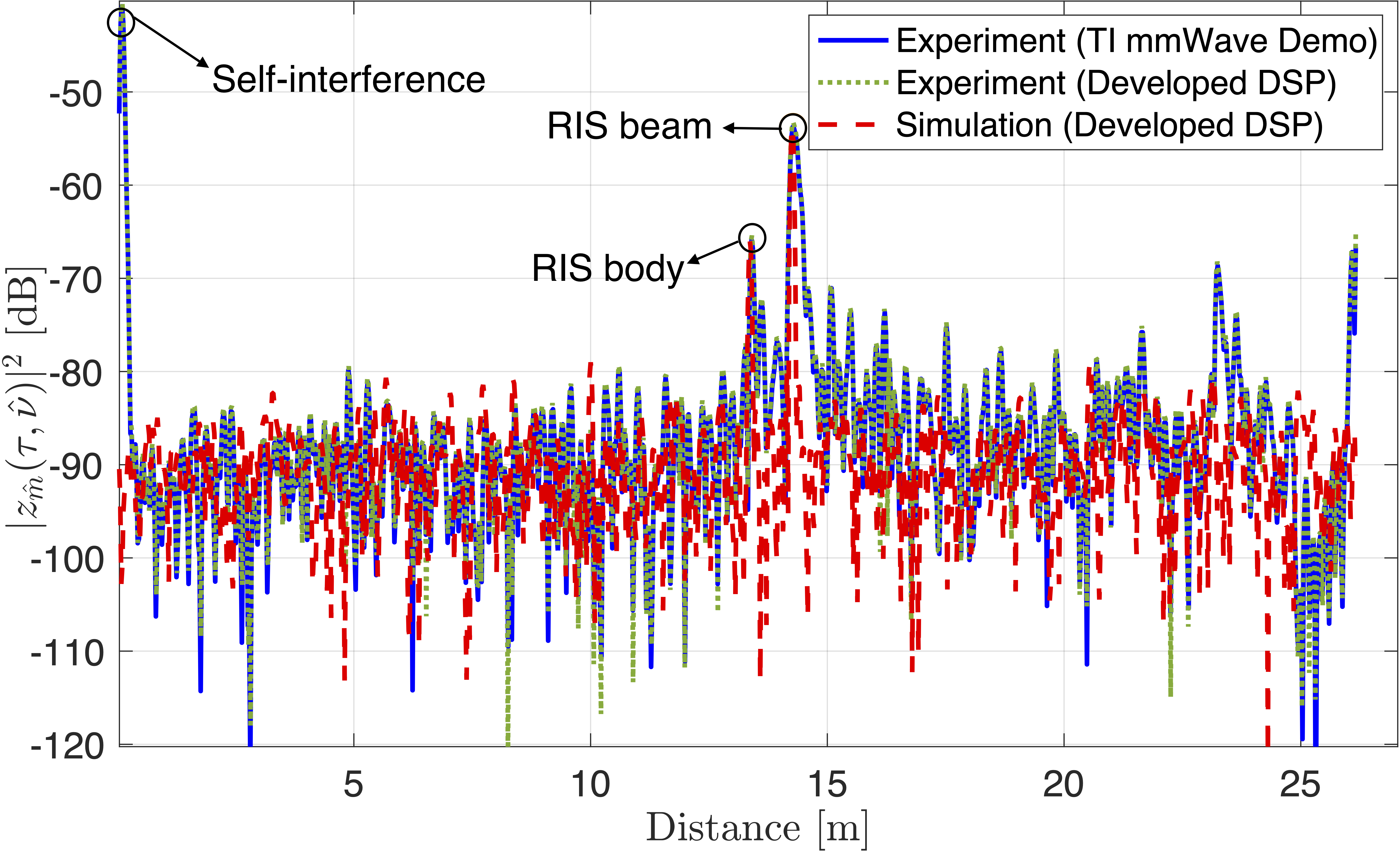}}    
    \par
    \caption{Comparison of 1D distance profile results with the experiment data by the \ac{ti} mmWave kit~(solid line) and by the proposed parameter estimation method~(dotted line), and simulation data by the designed signal model and proposed estimation method~(dashed line), at $\nu \approx 0$ and $\bm{\phi}=[0\degree,0\degree]^\top$ corresponding to the $\hat{m}$-th RIS sweep angle.}
    \label{Fig:Comparison}
\end{centering}
\end{figure}
    Utilizing the collected experimental data in the testbed, we investigate the 2D distance and velocity spectrum, obtained by~\eqref{eq:2DDFT} at the RIS sweep angle corresponding to $\bm{\phi}_m = [0,0]^\top$, shown in Fig.~\ref{Fig:RDM}.
    From the 2D spectrum, we see the first three largest peaks, 
    where the first is the self-interference due to the Tx antenna leakage to Rx in monostatic~(co-located) radars~\cite{Self-Interf},
    the second by the physical reflection by the RIS, and the third by the retransmitted signal at the RIS.
    Therefore, we filter out the first peak and then exploit the parameter estimation method~(Sec.~\ref{sec:GeoParaEst}).
    
    Fig.~\ref{Fig:AoD} shows the average power over the frames of~\eqref{eq:P_ave}, to estimate the AoD.  
    The true AoD is $\bm{\phi}=[0\degree,0\degree]^\top$ corresponds to the frames highlighted by the red bar, and estimated AoD $\hat{\bm{\phi}}=[-3\degree,0\degree]^\top$ corresponds to the frames marked by the red circle.
    The average power fluctuates sharply in the initial few hundred frames, then gradually increases, and eventually gradually decreases after passing about the midpoint of the frames. 
    This behavior is attributed to the RIS being turned on and off repeatedly during the initial few hundred frames for synchronization purposes, followed by the beam-steering sweep with the angular range from -45$\degree$ (starting at the 150-th frame) to 45$\degree$ in increments of 1.5$\degree$.

\vspace{-4mm}
    \subsection{Simulation Results and Extrapolation}

    \begin{figure}
    \centering
    \includegraphics[width=1\columnwidth]{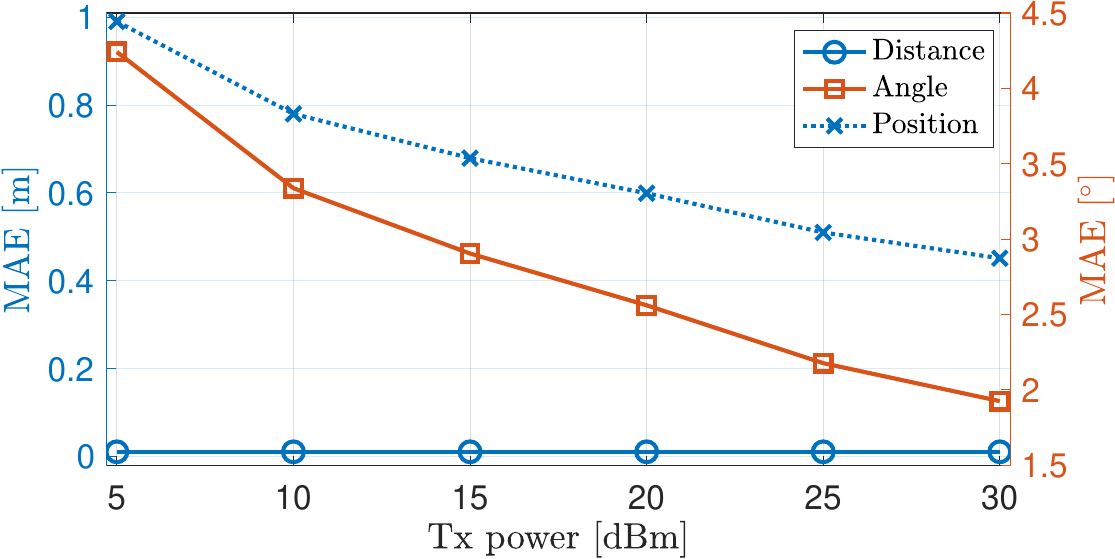}
    \caption{MAEs of distance, angle, and position estimates with the different Tx powers.}
    \label{Fig:Ptx}\vspace{-4mm}
\end{figure}

\begin{figure}
    \centering
    \includegraphics[width=1\columnwidth]{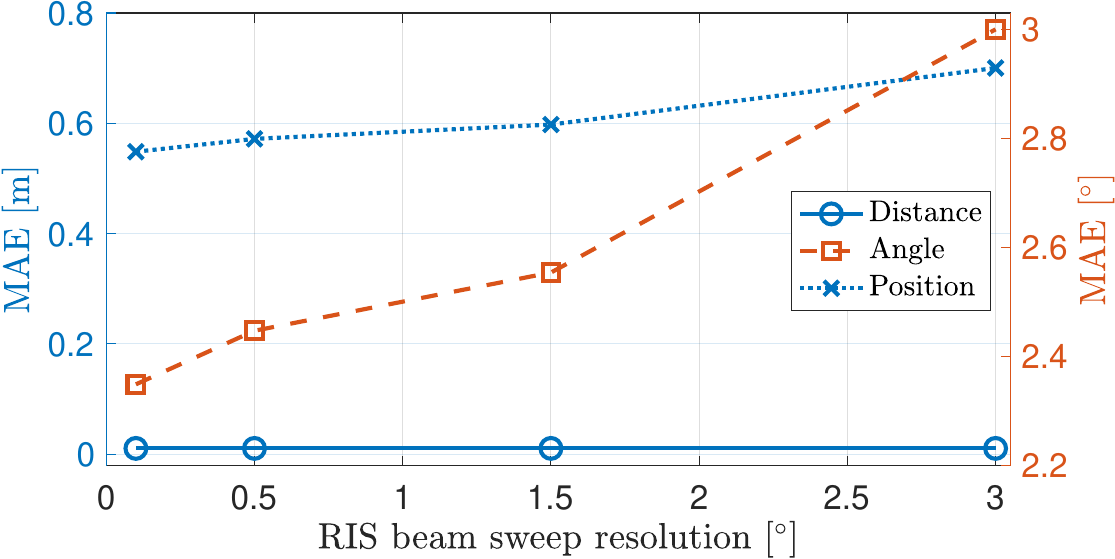}
    \caption{MAEs of distance, angle, and position estimates with the different RIS beam sweep resolutions.}
    \label{Fig:Beam}\vspace{-4mm}
\end{figure}
\begin{figure}[t!]
    \centering
    \includegraphics[width=1\columnwidth]{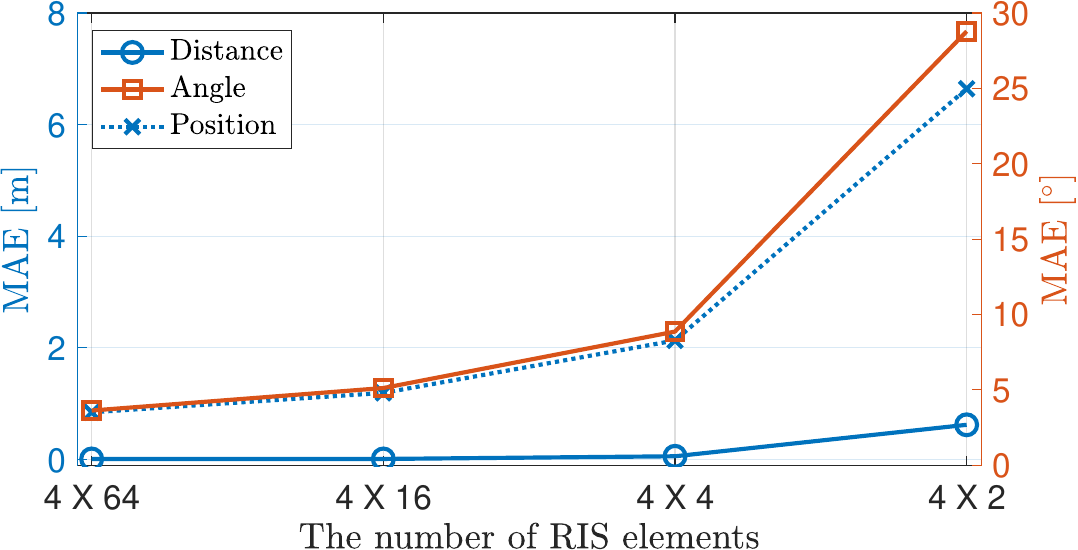}
    \caption{MAEs of distance, angle, and position estimates with the different number of RIS elements.}
    \label{Fig:NRIS}
\end{figure}
    To validate the developed \ac{dsp} unit using MATLAB including the formulated signal model~\eqref{eq:beat_signal} and the proposed geometric parameters estimation method~(Sec.~\ref{sec:GeoParaEst}), in Fig.~\ref{Fig:Comparison}, we compare the 1D distance profile results, at $\nu \approx 0$ and $\bm{\theta}=[0,0]^\top$. 
    The experiment data is analyzed by the \ac{ti} mmWave radar~(solid line) and by the proposed parameter estimation method~(dotted line), which is also applied to the simulation data~(dashed line).
    In the dashed line, the first peak that originated from the reflected signal from the UE itself is absent as this self-interference is not modeled in the simulated data.
    We demonstrate that the developed \ac{dsp} is valid as the solid line matches the dotted line, and the dashed line has the same two peaks corresponding to physical reflection at the RIS and retransmitted signal at the RIS, as the solid and dotted lines. 
    
    Figs.~\ref{Fig:Ptx}--\ref{Fig:NRIS} show the geometric parameters and positioning estimation errors with the different setup configurations.
    Fig.~\ref{Fig:Ptx} presents the results with the different Tx powers.
    The distance estimation error does not vary for the range of Tx power values for which the peaks of the spectrum can be resolved in the distance dimension with a fixed distance search grid~($N_\text{DFT}=1199$) on the distance profile.
    We see the smaller AOD error with the larger Tx power.
    Due to the smaller AOD error with the larger Tx power, while maintaining the same distance errors, the positioning error also decreases as the Tx power increases.
    Fig.~\ref{Fig:Beam} shows the results with various RIS beam sweep resolutions.
    The fine beam leads to a smaller AOD estimation error while maintaining the distance estimation error due to the resolvable peaks with a fixed distance search grid on the distance profiles.
    Therefore, the positioning error decreases with the fine RIS beam sweep resolution.
    Fig.~\ref{Fig:NRIS} presents the results with the different number of RIS elements.
    The distance estimation error also decreases with the larger number of RIS elements.
    This happens because the two peaks corresponding to physical reflection at the RIS and re-transmitted signal at the RIS are resolvable with the larger number of RIS elements.
    The AOD estimation error decreases as the number of RIS elements increases.
    Therefore, 
    the positioning error decreases with the large number of RIS elements.
    
\section{Conclusions}
    In this paper, we demonstrated the feasibility of \ac{ris}-enabled self-localization without AP intervention through both experimental and simulation results.
    In the experimental testbed, we assessed the developed \ac{dsp} unit by comparing the 
    1D distance profile with the \ac{ti} mmWave radar evaluation kit.
    We confirmed the validity of our designed signal model, which adheres to the deterministic model, in the signal propagation environment of the 60 GHz \ac{fmcw} radar.
    For future work, we plan to explore the feasibility of multiple target tracking and SLAM, with the assistance of RISs.
    \balance
\bibliographystyle{IEEEtran}
\bibliography{refs}
\end{document}